\documentclass[%
 reprint,
 amsmath,amssymb,
 aps,prc, twocolumn,
 tightenlines, superscriptaddress,
 nofootinbib
]{revtex4-1}

\usepackage{graphicx}
\usepackage{dcolumn}
\usepackage{bm}
\usepackage{natbib}
\usepackage{boldline}
\usepackage[T1]{fontenc}

\newcommand{\nuc}[2]{\hbox{$^{#1}$#2}}
\newcommand{\state}[3]{\hbox{$#1^{#2}_{#3}$}}
\newcommand{\pic}[2][1.0]{\includegraphics[width=#1\columnwidth]{#2}}
\newcommand{\picwide}[2][2.0]{\includegraphics[width=#1\columnwidth]{#2}}

\newcommand{\trans}[6]{\state{#1}{#2}{#3}~$\rightarrow$~\state{#4}{#5}{#6}}
\newcommand{\cascade}[9]{\state{#1}{#2}{#3}~$\rightarrow$~\state{#4}{#5}{#6}~$\rightarrow$~\state{#7}{#8}{#9}}
\newcommand{\transtentleft}[6]{(\state{#1}{#2}{#3})~$\rightarrow$~\state{#4}{#5}{#6}}

\newcommand{\mgcm}[0]{$~\text{mg}/\text{cm}^2~$}

\newcommand{\placedtrans}[0]{$15$~}
\newcommand{\newtrans}[0]{$29$~}

\begin{document}


\title{Probing the role of proton cross-shell excitations in \nuc{70}{Ni} using nucleon knockout reactions}

\author{B.\ Elman}
 \affiliation{National Superconducting Cyclotron Laboratory,
      Michigan State University, East Lansing, Michigan 48824, USA}
 \affiliation{Department of Physics and Astronomy,
      Michigan State University, East Lansing, Michigan 48824, USA}
\author{A.\ Gade}
    \affiliation{National Superconducting Cyclotron Laboratory,
        Michigan State University, East Lansing, Michigan 48824, USA}
    \affiliation{Department of Physics and Astronomy,
        Michigan State University, East Lansing, Michigan 48824, USA}
\author{R.\ V.\ F.\ Janssens}
 \affiliation{Department of Physics and Astronomy, University of North
      Carolina at Chapel Hill, Chapel Hill, North Carolina 27599, USA and
      Triangle Universities Nuclear Laboratory, Duke University, 
      Durham, North Carolina 27708, USA}  
\author{A.\ D.\ Ayangeakaa}
   \affiliation{United States Naval Academy, Annapolis, Maryland, 21402, USA}
\author{D.\ Bazin}
    \affiliation{National Superconducting Cyclotron Laboratory, Michigan State University, East Lansing, Michigan 48824, USA}
 \affiliation{Department of Physics and Astronomy,
      Michigan State University, East Lansing, Michigan 48824, USA}
\author{J.\ Belarge}
    \affiliation{National Superconducting Cyclotron Laboratory, Michigan State University, East Lansing, Michigan 48824, USA}
\author{P.\ C.\ Bender}
  \altaffiliation[Present Address: ]{Department of Physics, University of
  Massachusetts Lowell, Lowell, Massachusetts 01854, USA}
  \affiliation{National Superconducting Cyclotron Laboratory,
        Michigan State University, East Lansing, Michigan 48824, USA}
\author{B.\ A.\ Brown}
    \affiliation{National Superconducting Cyclotron Laboratory,
        Michigan State University, East Lansing, Michigan 48824, USA}
    \affiliation{Department of Physics and Astronomy,
        Michigan State University, East Lansing, Michigan 48824, USA}
\author{C.\ M.\ Campbell}
    \affiliation{Nuclear Science Division, Lawrence Berkeley National Laboratory, 
    Berkeley, California 94720, USA}
\author{M. P. Carpenter}
   \affiliation{Physics Division, Argonne National Laboratory,
   Argonne, Illinois 60439, USA}
\author{H.\ L.\ Crawford}
    \affiliation{Nuclear Science Division, Lawrence Berkeley National Laboratory, 
    Berkeley, California 94720, USA}
\author{B.\ P.\ Crider}
   \altaffiliation[Present Address: ]{Department of Physics and Astronomy,
   Mississippi State University, Mississippi State, Mississippi 39762, USA}
   \affiliation{National Superconducting Cyclotron Laboratory,
      Michigan State University, East Lansing, Michigan 48824, USA}
\author{P.\ Fallon}
    \affiliation{Nuclear Science Division, Lawrence Berkeley National Laboratory, 
    Berkeley, California 94720, USA}
\author{A.\ M.\ Forney}
   \affiliation{Department of Chemistry,
        University of Maryland, College Park, Maryland 20742, USA}
\author{J.\ Harker}
   \affiliation{Physics Division, Argonne National Laboratory,
   Argonne, Illinois 60439, USA}
   \affiliation{Department of Chemistry,
        University of Maryland, College Park, Maryland 20742, USA}

\author{S.\ N.\ Liddick}
   \affiliation{National Superconducting Cyclotron Laboratory,
      Michigan State University, East Lansing, Michigan 48824, USA}
 \affiliation{Department of Chemistry,
      Michigan State University, East Lansing, Michigan 48824, USA}
\author{B.\ Longfellow}
   \affiliation{National Superconducting Cyclotron Laboratory,
      Michigan State University, East Lansing, Michigan 48824, USA}
 \affiliation{Department of Physics and Astronomy,
      Michigan State University, East Lansing, Michigan 48824, USA}
\author{E.\ Lunderberg}
   \affiliation{National Superconducting Cyclotron Laboratory,
      Michigan State University, East Lansing, Michigan 48824, USA}
 \affiliation{Department of Physics and Astronomy,
      Michigan State University, East Lansing, Michigan 48824, USA}
\author{C.\ J.\ Prokop}
   \affiliation{National Superconducting Cyclotron Laboratory,
      Michigan State University, East Lansing, Michigan 48824, USA}
 \affiliation{Department of Chemistry,
      Michigan State University, East Lansing, Michigan 48824, USA}
\author{J.\ Sethi}
   \affiliation{Department of Chemistry,
        University of Maryland, College Park, Maryland 20742, USA}
\author{R.\ Taniuchi}
   \altaffiliation[Present Address: ]{Department of Physics, University of
   York, York YO10 5DD, United Kingdom}
\affiliation{Nuclear Science Division, Lawrence Berkeley National Laboratory, 
    Berkeley, California 94720, USA}
   \affiliation{Department of Physics, University of Tokyo, Hongo, Bunkyo, Tokyo 113-0033, Japan}
   \affiliation{RIKEN Nishina Center, 2-1 Hirosawa, Wako, Saitama 351-0198,
   Japan}
\author{W.\ B.\ Walters}
   \affiliation{Department of Chemistry,
        University of Maryland, College Park, Maryland 20742, USA}
\author{D.\ Weisshaar}
    \affiliation{National Superconducting Cyclotron Laboratory,
      Michigan State University, East Lansing, Michigan 48824, USA}
\author{S.\ Zhu}
   \affiliation{Physics Division, Argonne National Laboratory,
   Argonne, Illinois 60439, USA}

\date{\today}

\begin{abstract}
    The neutron-rich Ni isotopes have attracted attention in recent years due
    to the occurrence of shape or configuration coexistence. We report
    on the difference in population of excited final states in \nuc{70}{Ni}
    following $\gamma$-ray tagged one-proton, one-neutron, and two-proton
    knockout from \nuc{71}{Cu}, \nuc{71}{Ni}, and \nuc{72}{Zn} rare-isotope
    beams, respectively. Using variations observed in the relative transition
    intensities, signaling the changed population of specific final states in
    the different reactions, the role of neutron and proton configurations in
    excited states of \nuc{70}{Ni} is probed schematically, with the goal of
    identifying those that carry, as leading configuration, proton excitations
    across the $Z = 28$ shell closure. Such states are suggested in the
    literature to form a collective structure associated with prolate
    deformation. Adding to the body of knowledge for \nuc{70}{Ni}, \newtrans
    new transitions are reported, of which \placedtrans are placed in its
    level scheme.
\end{abstract}

\maketitle

\section{Introduction}
    Since its first production 30 years ago in neutron-induced fission of
    \nuc{235}U~\cite{armbruster1987}, the neutron-rich nucleus \nuc{70}Ni has
    been the subject of continued experimental and theoretical efforts due to its
    importance for guiding nuclear structure models along the benchmark
    proton-magic Ni isotopic chain~\cite{tsunoda2014, chiara2015, crider2016,
    morales2017, prokop2015, Wieland2018} as well as for low-entropy
    $r$-process nucleosynthesis contributing to the $A \sim 80$ abundance
    peak~\cite{Surman2008,Liddick2016,Spyrou2016,larsen2018}. From the
    nuclear structure perspective, \nuc{70}Ni displays a number of interesting
    phenomena such as the presence of low-lying electric dipole
    strength~\cite{Wieland2018} and of shape
    coexistence~\cite{tsunoda2014,crider2016,prokop2015}. Shape coexistence,
    indeed, appears to emerge as a common feature in the neutron-rich Ni
    isotopes, as evidenced by recent work on
    \nuc{66,68,70}{Ni}~\cite{suchyta2014, tsunoda2014, chiara2015, crider2016,
    prokop2015, leoni2017} and predictions for \nuc{78}{Ni}~\cite{Nowacki2016,
    taniuchi2019}.  The energetics and properties of the coexisting structures
    provide invaluable information on proton and neutron cross-shell
    excitations overcoming the relevant (sub)shell
    gaps~\cite{tsunoda2014,Nowacki2016,Gade2016}.
                                        
    The neutron-rich Ni isotopes highlight the drastic effects of shell
    evolution that can occur when adding or removing only a few nucleons within
    an isotopic chain. In \nuc{68}{Ni}, there are three known \state{0}{+}{} states
    associated with different shapes~\cite{suchyta2014,crider2016,tsunoda2014}:
    the spherical ground state, an oblate deformed level at 1604~keV, and a
    prolate deformed one at 2511~keV. In \nuc{70}{Ni}, the ground state is
    predicted to be slightly oblate~\cite{tsunoda2014} and a candidate
    for the expected prolate deformed (\state{0}{+}{2}) level has been reported
    recently from $\beta$-decay studies~\cite{prokop2015}. Monte Carlo shell-model 
    calculations by Tsunoda {\it et al.}~\cite{tsunoda2014} predicted
    this prolate minimum to be considerably deeper in \nuc{70}{Ni} than in
    \nuc{68}{Ni}. The proposed (\state{0}{+}{2}) level was tentatively
    established at 1567~keV~\cite{prokop2015}, indeed considerably below the
    proposed prolate state at 2511 keV excitation energy in
    \nuc{68}{Ni}~\cite{chiara2015,crider2016}.
    
    In general, it is interesting to explore the (band) structures built on top
    of shape-coexisting $0^+$ states as they may provide insights into the
    nature of the excitations involved; i.e., whether they are associated with
    deformation and, possibly, collective rotation. In \nuc{70}{Ni},
    shell-model calculations were only able to reproduce the proposed
    \state{2}{+}{2} state at 1868 keV, which was suggested to feed the $(0^+_2)$
    level~\cite{prokop2015}, when proton excitations across the $Z = 28$ shell
    closure were included in the model space~\cite{chiara2015}. The deformed
    structures in \nuc{70}{Ni} have also been discussed within the Nilsson
    scheme in Ref.~\cite{morales2017}. Through the $\beta$ feeding
    observed 
    in the \nuc{70}{Co} $\rightarrow$ \nuc{70}{Ni} decay, a connection was made
    between the prolate states suggested by the shell model in \nuc{70}{Ni} and
    the 
    deformed $1/2^-$ proton intruder state of \nuc{67}{Co} with a proposed
    $1/2^-$[321] Nilsson configuration~\cite{pauwels2008} and the \nuc{70}{Co}
    $(1^+,2^+)$ ground state, potentially arising from the coupling of this
    specific proton
    configuration with a $1/2^-$[301] neutron. The strong $\beta$ feeding of the
    $2^+_2$ state in \nuc{70}{Ni} from the \nuc{70}{Co} ground state
    was then conjectured to be an indicator that the two states carry a similar
    deformation~\cite{morales2017}.

Such emerging,
    likely deformed, intruder configurations may result in band  structures
    that would provide stringent tests for nuclear models as they require large
    configuration spaces and the inclusion of cross-shell excitations.
    Two-proton knockout from neutron-rich nuclei has been used in the past to
    selectively probe cross-shell proton excitations in proton-magic
    nuclei~\cite{gade2006}. 
    
    Here, results are reported from three different measurements that use
    complementary one-proton, one-neutron, and two-proton nucleon knockout
    reactions to populate excited states in \nuc{70}{Ni}. These
    different reactions enable the identification of dominant proton and
    neutron configurations in the wave functions of excited states. Although it
    is impossible to directly observe the de-excitation from the
    (\state{0}{+}{2}) level with in-beam $\gamma$-ray spectroscopy at 40\% of
    the speed of light, due to the state's long mean lifetime of $\tau(\state{0}{+}{2}) =
    2.38^{+0.43}_{-0.36}~\text{ns}$~\cite{crider2016}, evidence against some
    previously proposed candidates for the \trans{2}{+}{2}{0}{+}{2}
    transition~\cite{chiara2015} is provided. In addition, \placedtrans new
    transitions are placed in the level scheme, and the role of excitations
    across the $Z = 28$ gap in forming the predicted prolate deformed structure built
    on top of the \state{0}{+}{2} state of \nuc{70}{Ni} is explored. It is
    important to note that the suspected prolate shape in \nuc{70}{Ni} is
    interpreted as such based solely on the aforementioned shell-model
    calculations~\cite{tsunoda2014} where it is understood as resulting from a
    dominant proton particle-hole configuration. This shape cannot be directly
    inferred from the present data.
    
We note that the approach of exploiting complementary nucleon-subtracting or nucleon-adding direct reactions to disentangle the proton and neutron character and particle-hole content of final states has been used widely across the
nuclear chart, for example with $\gamma$-ray tagging, in the \nuc{208}{Pb} region~\cite{Sch97} and, most
recently, in the $N=20$ and $28$ islands of inversion~\cite{Pet15,Gade16} to probe shell evolution.

\section{Experiment}
    The  measurements were performed at the National Superconducting Cyclotron
    Laboratory~\cite{nsclref} at Michigan State University. The secondary beams
    of \nuc{71}{Cu}, \nuc{71}{Ni} and \nuc{72}{Zn} were produced from
    projectile fragmentation of a \nuc{76}{Ge} beam, accelerated by the K500
    and K1200 coupled cyclotrons to 130 MeV per nucleon. The primary beam
    impinged on a 399~mg/cm$^2$-thick \nuc{9}{Be} production target. An
    aluminum wedge degrader with an areal density of 300~mg/cm$^2$, located at
    the mid-acceptance position of the A1900 fragment
    separator~\cite{a1900ref}, was used to select the fragments of interest
    within the three different secondary beam cocktails. The identification of
    the secondary beam components of interest was accomplished using
    time-of-flight differences. The secondary beams interacted with the
    reaction target at energies of 80.2, 82.6, and
    76.5~MeV/nucleon for the \nuc{71}{Cu}, \nuc{71}{Ni}, \nuc{72}{Zn}
    projectiles, respectively.
   
    Two \nuc{9}{Be} reaction targets were used during this experiment: one with
    an areal density of 100\mgcm for the one-proton knockout reaction, and
    another of 188\mgcm thickness for the one-neutron and two-proton knockout
    reactions. For each setting, the target was located at the reaction target
    position of the S800 spectrograph~\cite{s800ref}. The event-by-event
    identification of the reaction residues and the trajectory
    reconstruction utilized the detection system of the spectrograph's focal
    plane, consisting of an ionization chamber, two xy-position-sensitive
    cathode-readout drift chambers (CRDCs), and a plastic timing scintillator
    that also served as the particle trigger~\cite{s800fpref}. 
    
\begin{figure}[ht!]
         \begin{center}
             \pic{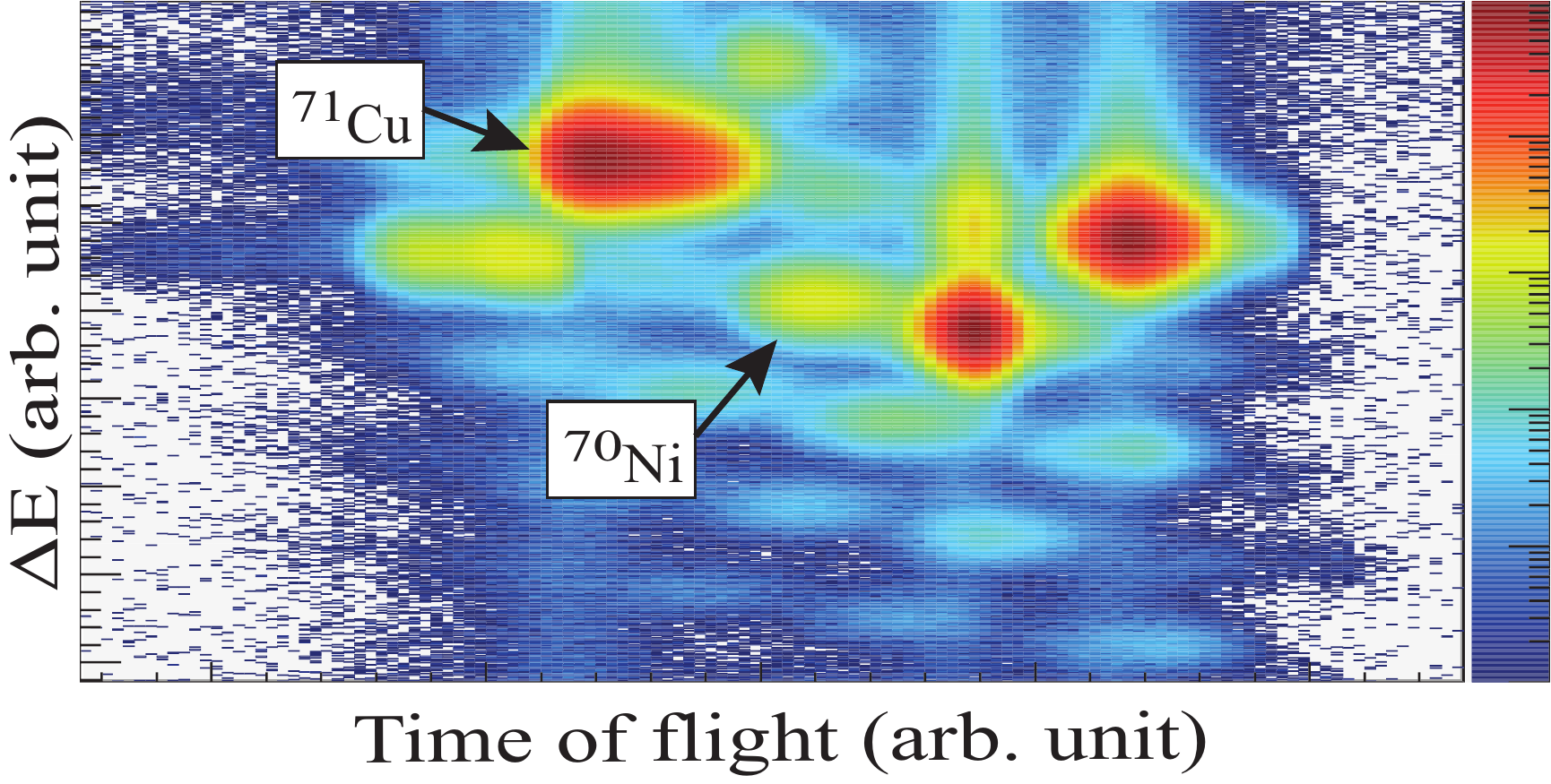}
             \caption{ 
                 (Color online) Particle identification plot (energy loss
                 vs. time of flight) for the setting centered on the one-proton
                 knockout from \nuc{71}{Cu}. The beam components and different
                 reaction products are identifiable and were cleanly separated
                 with additional software gates on angles and position
                 information in the focal plane.
             }
             \label{fig:pid}
         \end{center}
     \end{figure}

    An example of the identification of the reaction products emerging from the
    \nuc{9}{Be} target for the \nuc{71}{Cu} one-proton knockout setting is
    given in Fig.~\ref{fig:pid}, where the energy loss measured with the S800
    ionization chamber is displayed versus the ion's trajectory-corrected time
    of flight measured between two plastic scintillators. The \nuc{70}{Ni}
    knockout residues can be separated from the other reaction products;
    primarily Zn, Cu, Ni, Co, and Fe isotopes. Additional gates on angles and
    positions in the focal plane were used to remove any contamination by the
    tails of neighboring nuclei. The identification of the \nuc{70}{Ni}
    residues in the one-neutron and two-proton knockout settings proceeded in
    the same way. 

    The Be reaction target was surrounded by the Gamma-Ray Energy Tracking In-beam
    Nuclear Array (GRETINA)~\cite{paschalis2013}, an array of 36-fold
    segmented high-purity germanium crystals used for in-flight $\gamma$-ray
    detection. At the time of the experiment, the array was composed of nine
    modules that housed four detectors each in a common cryostat. GRETINA's
    spatial resolution through signal decomposition provided event-by-event
    Doppler-reconstruction capability for $\gamma$ rays emitted
    in-flight~\cite{dirk2017}. This Doppler correction takes into account the
    reconstructed trajectory and kinetic energy of each particle at the S800
    target position, in addition to the $\gamma$-ray interaction points provided
    by GRETINA. 

    After passing the scintillator used for the particle trigger, the beam-like
    reaction residues were implanted in an aluminum plate in front of a CsI(Na)
    hodoscope array~\cite{meierbachtol2011}, which was arranged in the IsoTagger
    configuration~\cite{wimmer2015}. This configuration enabled the
    identification of long-lived states with lifetimes between 100 ns and
    several ms, such as the 2861-keV level ($\tau = 335(1)~\text{ns}$~\cite{NNDC}) 
    in \nuc{70}{Ni}.

    Both \nuc{71}{Cu} and \nuc{71}{Ni} projectiles exhibit isomers in their
    level schemes that affect the population of excited states in \nuc{70}{Ni},
    if present in the beam. In contrast, \nuc{72}{Zn} has no known long-lived
    state. The isomeric content of the beams was measured by placing a
    $5.1~\text{mm}$-thick Al stopper at the target position of the S800
    spectrograph and measuring the presence of isomeric states with GRETINA
    through their characteristic $\gamma$-ray transitions. In the case of
    \nuc{71}{Cu}, there is a known (\state{19/2}{-}{ }) state at 2756~keV with
    a mean lifetime of $\tau=391(20)~\text{ns}$~\cite{NNDC}. The isomeric
    content in the \nuc{71}{Cu} beam was determined to be $0.47(7)\%$, based on
    the detection of the 133-keV $\gamma$ ray from this state.
    
    In the case of \nuc{71}{Ni}, estimating the isomeric content is more
    challenging because the long-lived state does not decay directly by $\gamma$-ray
    emission. Instead, the (\state{1/2}{-}{ }) state at 499~keV with a mean
    lifetime of $\tau=3.3(4)~\text{s}$~\cite{NNDC} in \nuc{71}{Ni} undergoes
    $\beta$ decay into either the \state{3/2}{(-)}{ } ground state or the
    454-keV (\state{1/2}{-}{ }) excited level of
    \nuc{71}{Cu}~\cite{stefanescu2009}. At present, only a $40\%$ upper limit
    is available for the adopted value for the branching ratio to this 454-keV
    state in $\beta$ decay~\cite{NNDC}. As a result, based on the intensity of
    the 454-keV \transtentleft{1/2}{-}{ }{3/2}{(-)}{ } transition, a limit of
    $I_c > 6\%$ was derived for the \nuc{71}{Ni} isomeric content.
    
    \section{Methodology and results}
      \begin{table*}[]
        \caption{Measured intensities for all observed \nuc{70}{Ni} transitions
          relative to the number of detected \nuc{70}{Ni} residues in the
          specific reaction channel expressed as a percentage. The error
          includes both statistical uncertainties and systematic contributions
          from varying the fit assumption for peak shapes and backgrounds, and
          a $2.1\%$ uncertainty from the efficiency determination. Note that no
          feeding subtraction is included here.} 
          \begin{ruledtabular}
            \begin{tabular} {@{}ccccc@{}}
              State& Transition & Relative Intensities (\%) &  Relative Intensities (\%) & Relative Intensities (\%) \\
              Energy (keV) & Energy (keV) & 1p Knockout & 1n Knockout & 2p Knockout \\
              \hline
              1260(2) & 1260(2) & 44(2)  & 14.5(8) & 21(1)\\
              1868(2) & 609(2)  & 3.6(4) & 1.9(3) & 2.4(4) \\
                      & 1868(2) & 3.5(6) & 1.3(5) & 2.5(5) \\
              2230(3)    & 970(2)  & 16.5(9) & 5.7(7) & 7.8(8)  \\
              2509(4)    & 279(3)~\footnote{The expected intensity for the 279-keV
            transition is below the detection sensitivity in the one-neutron and
            two-proton reaction channels, so the intensity for these
            reaction channels was estimated based on the branching ratio
            measured in one-proton knockout.}  & 0.5(2) & 0.3(2)& 0.5(2) \\ 
                    & 640(2)  & 2.2(3) & 1.2(6) & 2.2(4) \\
                    & 1249(3)~\footnote{The discrepancy between the different
            measurements of the 1249-keV transition intensity relative to the
            640-keV one is due both to the difficulty of determining the
            intensity in a self-coincident doublet and to possible
            lifetime effects as discussed in Section~\ref{subsec:nko}. The measured intensities
            are too small to observe this transition in coincidence with the
            1260-keV $\gamma$ ray in the one-neutron and two-proton reactions.} & 2.4(4) &
            1.7(3) & 1.0(6) \\
              2912(4) & 234(4)~\footnote{The intensities given for the 234-keV
            $\gamma$ ray for the one-proton and two-proton knockout are
            determined based on the NNDC adopted branching ratio~\cite{NNDC}.}&
            0.4(1) & 1.5(4) & 0.6(2) \\ 
                  & 682(3)~\footnote{The splitting between the self-coincident 676-682 keV doublet is
              determined based on add-back coincidences for the one-proton
              knockout. This splitting was then used for the two-proton
            knockout as well.}  & 1.1(3)  & 3.9(6) & 1.6(5) \\ 
              2942(4) & 1682(3) & 1.3(3) & -       & 1.0(9) \\ 
              3215(4) & 1955(3) & 2.1(2) & 0.8(3) & 0.7(5)\\
              3297(5) & 385(3) & 1.3(2) & 2.6(7) & 1.3(5)\\
              3551(5) & 1321(4) & 0.7(1) & -       & - \\
              3588(6) & 676(4)  & 0.12(3) & -      & 0.16(5) \\
              3662(4) & 2402(3) & 0.8(2) & -       & -\\
                      & 1432(4) & 0.3(2) & -       & -\\
              3814(3) & 1584(2) & 4.4(4) & -       & -\\
              3846(5) & 3846(5)~\footnote{The peak is a little wider than expected and the energy was deduced based on the assumption of a single peak; the resulting energy is in agreement with~\cite{prokopthesis}.}  & 0.7(1) & -       & - \\
              3961(4) & 2701(3) & 1.6(3) & -       & - \\
              4017(4) & 1787(3) & 1.8(6) & -       & - \\
                      & 2757(3) & 1.4(2) & -       & - \\
              4293(4) & 479(3)  & 0.7(2) & -       & - \\
                      & 631(4)  & 0.9(2) & -       & -\\
                      & 2063(3) & 0.75(16) & -       & -\\
                      & 3033(5) & 1.2(2) & -       & -\\
              Unplaced & 424(5)  & -      & 0.6(3)& 0.7(3)\\
                       & 660(3)  & 0.6(4)& 1.2(11)& 1.2(4)\\
                       & 714(4)  & -  & - & 1.4(9) \\
                       & 915(5)  & 0.9(3)& 1.5(4)& - \\
                       & 930(5)  & 0.33(25)& 0.9(3)& -\\
                       & 958(5)  & 0.6(3)& 1.8(5)& -\\
                       & 1212(5) & -      & -      & 1.1(5)\\
                       & 1225(5) & -      & 1.7(4)& -\\
                       & 1440(3) & 0.8(3)& -      & -\\
                       & 1467(3) & 0.5(2)& -      & -\\
                       & 2026(3) & 1.2(2)& -      & 2.2(5)\\
                       & 2114(3) & 0.8(2)& 0.6(3)& -\\
                       & 2342(5) & 0.5(2)& -      & -\\
                       & 2980(4) & 0.4(3)& -      & -\\
            \end{tabular}
          \end{ruledtabular}
          \label{tab:res} \end{table*}

    In the direct one- or two-nucleon nucleon knockout reactions used here, the
    projectile of interest impinges on a \nuc{9}{Be} target, and one or two
    nucleons are removed leaving the projectile-like residue as a spectator to
    the sudden collision~\cite{Hansen2003}. These reactions are often used to
    quantify spectroscopic strength and probe shell-model spectroscopic factors
    or two-nucleon amplitudes. However, due to knockout from isomers that
    cannot be quantified event-by-event, as well as to (1) mostly unknown or
    tentative final-state quantum numbers, to (2)  complex associated
    configurations, and to (3) expected high level densities (see for
    example~\cite{Recchia2016}), the experiment was optimized for in-beam
    $\gamma$-ray spectroscopy~\cite{Gade2008}. Hence, $\gamma$-ray intensities
    per reaction residue will be compared across the different reaction
    channels to assess changes in population of final states rather than to
    obtain absolute cross sections. 

    Intensities of emitted $\gamma$ rays were determined for all three
    reactions. Uncertainties on these intensities are composed of
    statistical ones, as well as contributions from the efficiency
    determination ($2.1\%$)~\cite{dirk2017} and a specific uncertainty for each
    separate peak determined from fitting the spectra while varying the fit
    template and background model. A systematic energy uncertainty of 2.2~keV
    was determined by comparing observed peak energies with those adopted from
    the NNDC database~\cite{NNDC}. It was added in quadrature with the uncertainty
    determined from fitting the peaks. Due to the position sensitivity of
    GRETINA and the emission of $\gamma$ rays in-flight, excited-state
    lifetimes on the order of $10~\text{ps}$ or longer will displace peaks to
    lower Doppler-reconstructed energies than their actual value.  This is not
    reflected in the energy uncertainties quoted here.
    
\begin{figure}[h!]
      \begin{center}
        \pic{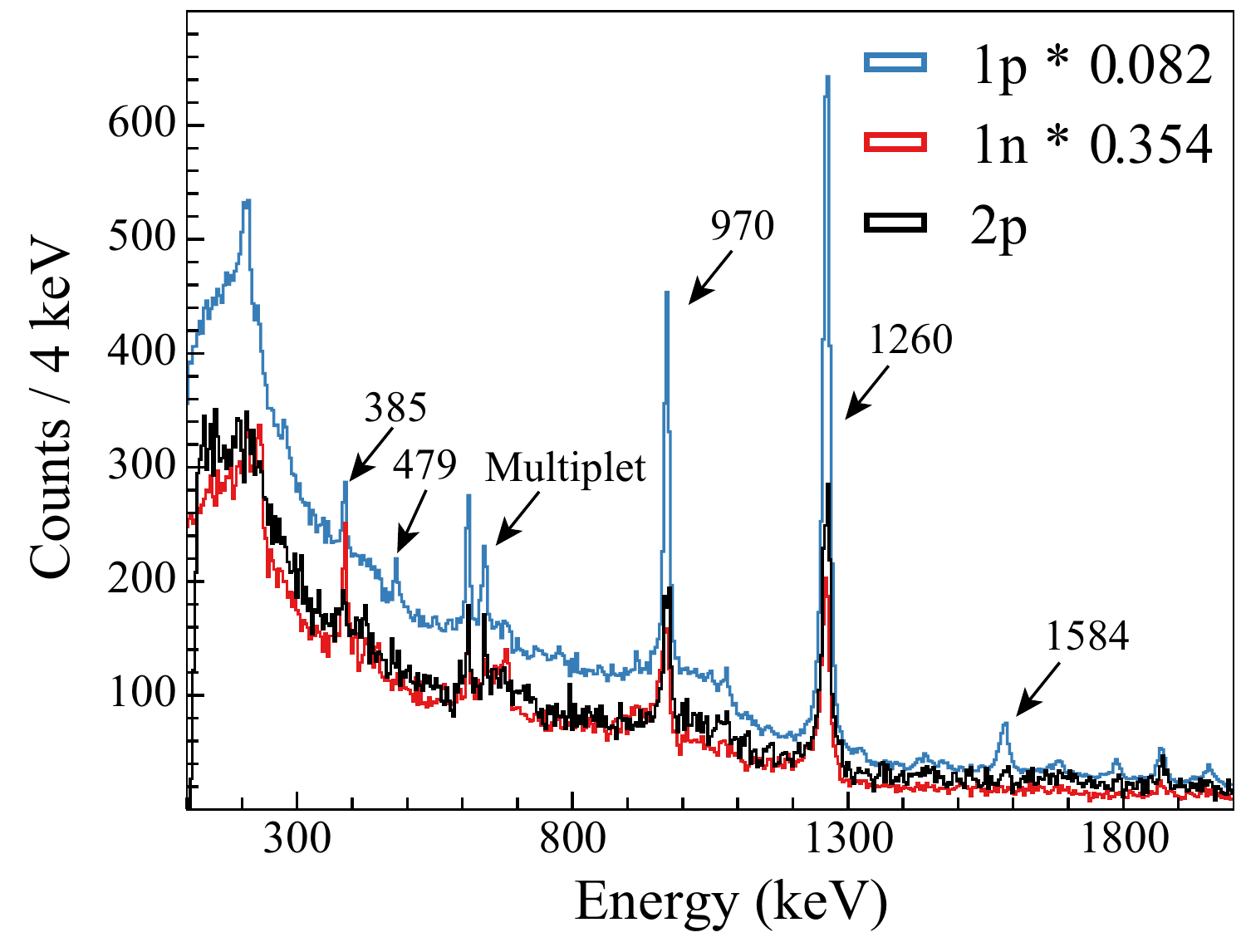}
        \caption{ 
          (Color online) Comparison of spectra from all three knockout
          reactions. The one-proton and one-neutron knockout spectra are
          scaled down by the ratio of the number of \nuc{70}{Ni} residues in
          the focal plane for the specific reaction channel to those detected in the
          lower statistics two-proton knockout. These scaling factors are
          given in the figure. 
        }
        \label{fig:allspec_leg}
      \end{center}
    \end{figure}

    The energy-dependent photopeak efficiency of the setup was
    determined~\cite{dirk2017} using standard calibration sources. For the
    in-beam response of the array, the Lorentz boost of the emitted
    $\gamma$-ray distribution was taken into account via GEANT4
    simulations~\cite{ucgretina}. All intensities were determined using
    $\gamma$-ray singles spectra (see Fig.~\ref{fig:allspec_leg}). Levels were
    placed into the level scheme based on their $\gamma\gamma$ coincidence
    relationships, taking advantage of nearest-neighbor addback routines for
    GRETINA~\cite{dirk2017}. 

\begin{figure*}[ht!]
      \begin{center}
        \picwide{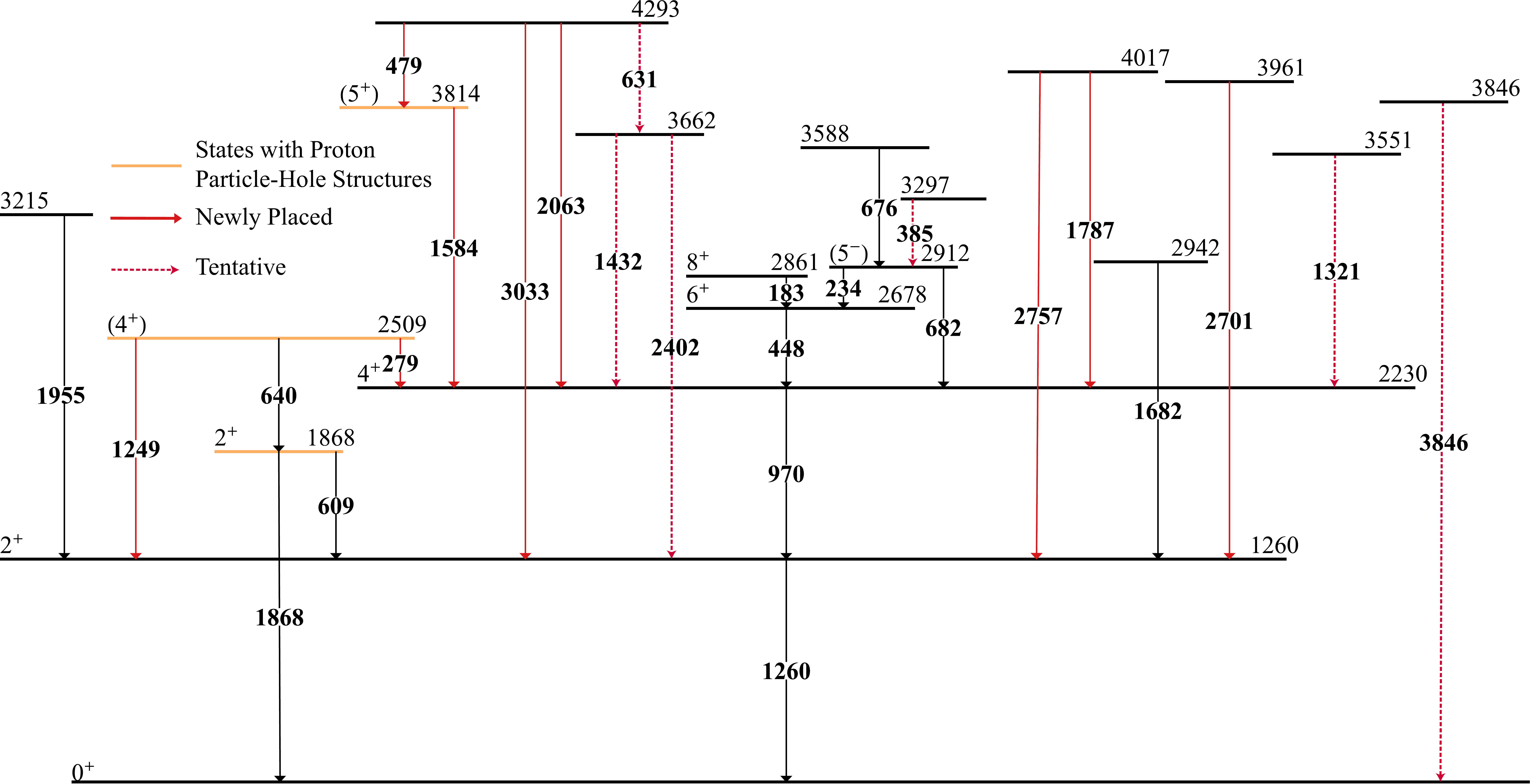}
        \caption{ 
          (Color online) Proposed \nuc{70}{Ni} level scheme containing all
          the transitions observed  in the present measurements. Red arrows indicate newly-placed transitions,
          and dashed red ones mark tentative placements. The states highlighted in
          orange are those associated with proposed proton particle-hole configurations
          (see text).
        }
        \label{fig:lvlscheme}
      \end{center}
    \end{figure*}

    The results for the measured intensities relative to the number of
    \nuc{70}{Ni} residues detected for each reaction channel are summarized in
    Table~\ref{tab:res}. No feeding subtraction was applied to the quoted
    intensities, in contrast to the feeding-subtracted state populations 
    discussed in Section~\ref{sec:disc}. Note that the fit to extract
    the intensities included all observed peaks as well as a background model,
    consisting of both an exponential component (used to model the prompt,
    beam-correlated background) and stopped lines from annihilation
    radiation and inelastic reactions of neutrons and other light particles
    within the germanium crystals or the aluminum beam pipe~\cite{loelius2016}.
    The results for placements in the level scheme are given in
    Fig.~\ref{fig:lvlscheme}. Twenty-nine new transitions were observed, of which
    \placedtrans were placed within the level scheme.

In Fig.~\ref{fig:allspec_leg}, the spectra from the one-proton and one-neutron
knockout reactions are normalized relative to the two-proton one (see caption
for details). The spectra from the three knockout reactions clearly display
differences in the population of excited final states. The largest differences
are the intensity of the 385-keV transition, whose relative intensity (see
Table~\ref{tab:res}) in one-neutron knockout is a factor of two larger than in
proton knockout, a number of transitions which appear uniquely in proton
knockout, such as the 1584- and 479-keV $\gamma$ rays, and the shape of the
multiplet between 600 and 700 keV, which varies in all the reaction channels,
indicating that some components in the structure change intensity, depending on
the reaction. The wide feature near 200 keV that appeared only in the one-proton
knockout exhibited no clear coincidence relationships with other transitions,
likely because it is in a region of the spectrum where the peak-to-background
ratio is poor. Nevertheless, it seems likely that there is more than one
transition in the region between 100 and 200 keV, forming the structure visible
in Fig.~\ref{fig:allspec_leg}. We note that, due to the high detection
efficiency at low $\gamma$-ray energies, the intensities of such transitions on
top of the high background would actually be small.   

In the following sub-sections, each reaction is discussed separately with a main
focus on differences between the three channels. For the discussions of
single-particle configurations, we remind the reader that (i) one-proton knockout from
the dominant $f_{7/2}$ or $p_{3/2}$ orbitals of \nuc{71}{Cu} populates
positive-parity 
final states in \nuc{70}{Ni} (the ground-state neutron occupations for
\nuc{71}{Cu} in the shell model with the jj44 interaction are 8.0, 0.06, 0.87,
0.05, and 0.03 for the $f_{7/2}$, $f_{5/2}$, $p_{3/2}$, $p_{1/2}$, and $g_{9/2}$
states, respectively), with negative-parity states originating from the
knockout of a $sd$-shell proton expected only at about 7 MeV~\cite{Mairle1969}
and that (ii) one-neutron knockout from the $f_{5/2}$, $p_{3/2}$, and $ p_{1/2}$
orbitals in \nuc{71}{Ni} leads to positive-parity states in \nuc{70}{Ni}, while
removal of a $g_{9/2}$ neutron populates negative-parity states. 

\subsection{One-proton knockout from \nuc{71}{Cu}}
    \label{subsec:pko}
\begin{figure}[ht]
      \begin{center}
        \pic{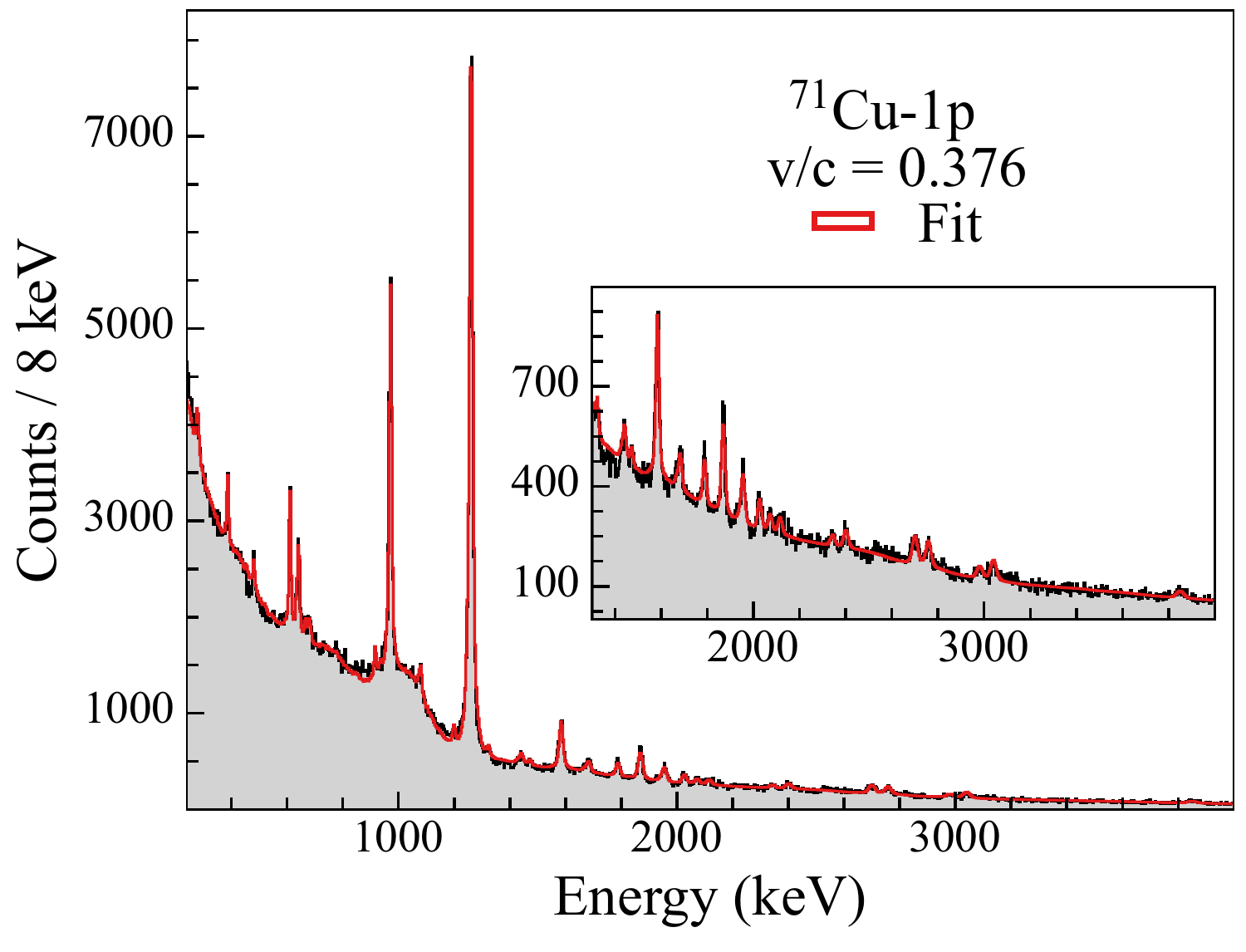}
        \caption{ 
          (Color online) Fit to the full spectrum of \nuc{70}{Ni} obtained in
          one-proton knockout from \nuc{71}{Cu}. This fit includes the
          simulated response function for all the peaks in the spectrum and a
          background composed of a double exponential to model the
          beam-correlated background in addition to stopped background lines
          from annihilation radiation and inelastic reactions of neutrons and
          other light particles on the beam pipe or the Ge detectors.
        }
        \label{fig:fitcu71}
      \end{center}
    \end{figure}

    In the one-proton knockout reaction, states in the yrast band and in the
    proposed prolate structure were populated more strongly than other levels.
    Within the latter structure, two new transitions were observed de-exciting
    the (\state{4}{+}{2}) state at 2509-keV to the yrast sequence. In addition,
    several higher-lying levels of undetermined spin-parity were populated.
    These decay primarily into the yrast states.  A fit to the
    spectrum for this reaction channel is presented in Fig.~\ref{fig:fitcu71}. 

    \begin{figure}[ht]
      \begin{center}
        \pic{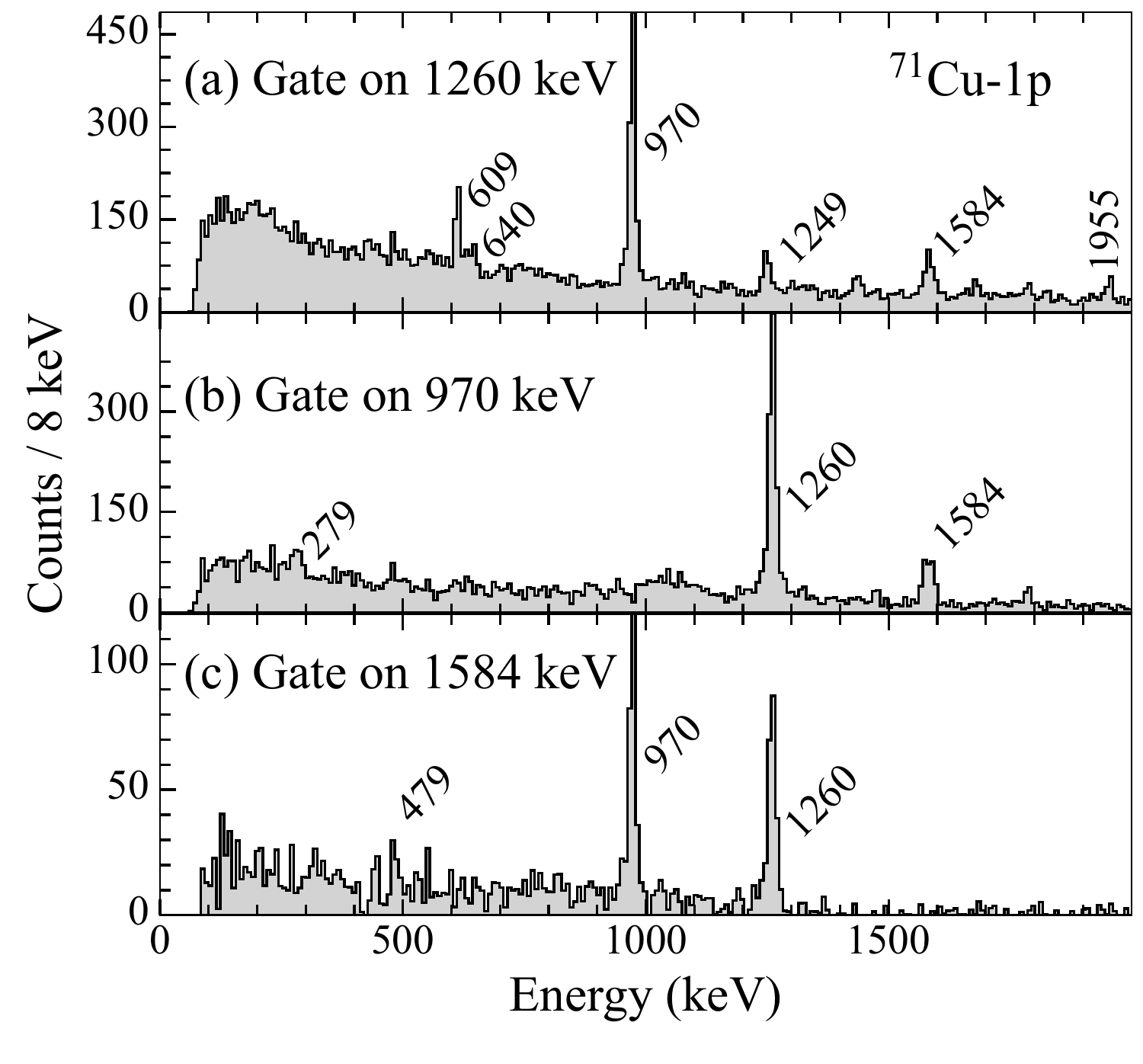}
        \caption{ 
          Coincidence spectrum used to place the 1584-keV transition in the
          one-proton knockout reaction. Clear coincidence relationships between the
          1260-970-1584 keV transitions allow placing this transition into
          the level scheme. Also shown in (a) and (b) are the two newly-placed
          transitions, 1249 and 279 keV, in what might be the prolate structure.
        }
        \label{fig:coinc_cu71}
      \end{center}
    \end{figure}

    Twenty-five previously unplaced transitions were observed in this reaction
    channel, and  \placedtrans of these were placed in the level scheme of
    Fig.~\ref{fig:lvlscheme}. Some transitions were difficult or impossible to
    place due to a combination of low statistics, lack of clear coincidence
    relationships with known peaks, or insufficient resolution in regions with
    multiple $\gamma$ rays close in energy. New transitions originating from
    the proposed prolate structure are presented in Fig.~\ref{fig:coinc_cu71}.
    Specifically, the \transtentleft{4}{+}{2}{2}{+}{1}, 1249-keV $\gamma$ ray
    is clearly seen in Fig.~\ref{fig:coinc_cu71}(a), while the 279-keV,
    \transtentleft{4}{+}{2}{4}{+}{1} transition is present in panel (b). The
    spin-parity quantum numbers for this (\state{4}{+}{2}), 2509-keV level are
    tentative as they are based on the observed coincidence relationships with
    known transitions from the \state{2}{+}{2} state as well as on the absence
    of the 640-keV, \transtentleft{4}{+}{2}{2}{+}{2} $\gamma$ ray in
    $\beta$-decay measurements~\cite{sawicka2003, mueller2000}. It should be
    noted that the two new 1249- and 279-keV transitions were not observed
    in the two-neutron knockout measurement of Ref.~\cite{chiara2015}. This is
    possibly due to the former being part of a doublet peak and the latter
    belonging to a region of the spectrum where the peak-to-background ratio
    was poor.

    The majority of newly placed transitions originate from higher-lying states
    that decay toward the established yrast levels. The strongest of these is
    the 1584-keV line feeding the \state{4}{+}{1}, 2230-keV state. Coincidence
    spectra, instrumental in placing this $\gamma$ ray, are found in
    Fig.~\ref{fig:coinc_cu71} where panel (a) provides a spectrum coincident
    with the \trans{2}{+}{1}{0}{+}{1} ground-state transition in which the
    1584-keV $\gamma$ ray is evident. The latter spectrum also displays
    transitions associated with the proposed prolate structure such as the
    newly placed 1249-keV $\gamma$ ray and the previously observed 609- and
    640-keV ones.  The spectrum in coincidence with the 970-keV transition of
    Fig.~\ref{fig:coinc_cu71}(b)  documents the placement of the 1584-keV
    $\gamma$ ray as feeding the \state{4}{+}{1} state from a newly placed
    3814-keV level. A placement as populating the \state{6}{+}{1} yrast state
    is ruled out, based on the $1.51(4)$~ns lifetime~\cite{mach2003} of the
    latter level prohibiting the observation of coincidence relationships for
    these fast-moving reaction products. Finally, the spectrum gated on the
    1584-keV line itself (Fig.~\ref{fig:coinc_cu71}(c)) demonstrates the
    expected presence of the 970-1260 keV, \cascade{4}{+}{1}{2}{+}{1}{0}{+}{1}
    cascade.  The observation of a weak 479-keV $\gamma$ ray is noteworthy as
    well, as discussed further below.
    
    As seen in Fig.~\ref{fig:lvlscheme}, the 3814-keV state appears to decay 
    only to the \state{4}{+}{1} state. Because of the large intensity of the 1584-keV
    transition feeding the \state{4}{+}{1} level, and the available orbitals from
    which protons can be removed, it is possible that the 3814-keV level is a
    (\state{5}{+}{1}) state with a $\pi1p_{3/2}^{+1}0f_{7/2}^{-1}$
    particle-hole configuration. However, these configuration and spin-parity
    assignments remain tentative.
    
    The 4017- and 4293-keV states are two newly-proposed levels that are
    populated as strongly as the 3814-keV one. As seen in
    Fig.~\ref{fig:lvlscheme}, both exhibit considerably more branching to
    other states when compared to the 3814-keV state. These decay branches include 
    intense, direct transitions to the \state{2}{+}{1} level, which strongly suggests
    that these states have spin $I \leq 4$.

    The 4293-keV state's strongest decay branch is to the \state{2}{+}{1}
    state, although there are three somewhat weaker transitions to the
    (\state{5}{+}{}) level (via the aforementioned 479-keV transition), the
    \state{4}{+}{1} one, and a new state at 3662 keV. This suggests a possible
    spin-parity assignment of (\state{3}{+}{}, \state{4}{+}{}).  Similar arguments hold
    for the 3662- and 4017-keV states, but because these have no connection to
    the 3814-keV, (\state{5}{+}{}) level, they can possibly correspond to
    (\state{2}{+}{}, \state{3}{+}{}, \state{4}{+}{}) states. These are
    most likely of positive parity, considering the available orbitals from which
    protons can be knocked out. Strong population of negative-parity states
    would require either indirect, possibly unresolved, feeding from
    positive-parity levels or knockout of protons from the $sd$ shell. Levels
    populated from the knockout of $sd$-shell protons are only expected to appear
    starting at energies near 7~MeV in the level scheme~\cite{Mairle1969}.

    The 1682-keV $\gamma$ ray reported here possibly corresponds to the
    1676-keV transition observed in two recent $\beta$-decay
    experiments~\cite{morales2017, prokop2015}. It was seen following the
    $\beta$ decay of the (\state{1}{+}{}, \state{2}{+}{}) state of
    \nuc{70}{Co}, and is observed here in both of the proton reaction channels,
    albeit only weakly in two-proton knockout. The reason for the potential
    discrepancy between the measured centroids in the different measurements is
    unclear.

    The 1955-keV transition only displays a clear coincidence with the
    1260-keV one (see Fig.~\ref{fig:coinc_cu71}(a)). This decay out of the
    3215-keV level appears in all reaction channels and has also
    been reported following $\beta$ decay~\cite{prokop2015, morales2017} from
    the (\state{1}{+}{}, \state{2}{+}{}) state of \nuc{70}{Co} as well as
    two-neutron knockout~\cite{chiara2015}. 

    The 3846-keV $\gamma$ ray was reported previously following the $\beta$
    decay of the low-spin isomer in \nuc{70}{Co}, and tentatively placed as
    feeding the proposed prolate 1868-keV state~\cite{prokopthesis}. This
    placement cannot be confirmed through coincidence relationships in the
    present data.  Due to the absence of coincidence information, it is
    tentatively placed here as directly feeding the ground state. This
    transition was observed almost exclusively in events with a detector
    multiplicity of 1, herewith supporting this placement. However, the fact
    that it could decay into the isomeric (\state{0}{+}{2}) state at 1567
    keV~\cite{prokop2015} cannot be ruled out in view of the associated
    lifetime.  It is unlikely to decay into the other known isomers; e.g., the
    \state{6}{+}{1} or \state{8}{+}{1} yrast levels, as it was observed in the
    $\beta$ decay from a (\state{1}{+}{}, \state{2}{+}{}) low-spin state.

    As can be seen in the level scheme of Fig.~\ref{fig:lvlscheme}, a 682-keV
    $\gamma$ ray is placed as de-exciting the proposed~\cite{chiara2015}
    2912-keV, (\state{5}{-}{}) level into the \state{4}{+}{1}, 2230-keV state. 
    In the present experiment, this transition forms an unresolved doublet with
    a $\gamma$ ray with a 676-keV approximate energy. The latter has been
    previously observed following the $\beta$ decay of the short-lived 
    (\state{6}{-}{}, \state{7}{-}{}) state in \nuc{70}{Co}~\cite{prokop2015}.
    The data indicate that the two components of the doublet are in mutual
    coincidence and, combining the observations from the literature with those
    from the present data, leads to the placements proposed in
    Fig.~\ref{fig:lvlscheme}. It should be noted that the direct population of
    the (\state{5}{-}{}) level would require proton removal from the rather
    deeply-bound $sd$ shell. As already mentioned, this is viewed as unlikely
    and, hence, this 2912-keV state is probably fed from higher-lying levels.

    As discussed above, the observation of $\gamma$ decay from
    negative-parity states in the proton-knockout reactions is expected to
    be the result of feeding by transitions from positive-parity states or
    high-lying, negative-parity levels populated by proton removal 
    from the $sd$ shell. Although the spin-parities of the states which appear
    coincident with the 234- and 682-keV transitions are unknown, the full
    intensity of the transitions from the (\state{5}{-}{ }) level in the
    proton-knockout channels can be accounted for through feeding of the
    coincident 676- and 385-keV transitions. The placement of the latter is
    discussed in the following section.

    \subsection{One-neutron knockout from \nuc{71}{Ni}}
    \label{subsec:nko}
\begin{figure}[ht!]
      \begin{center}
        \pic{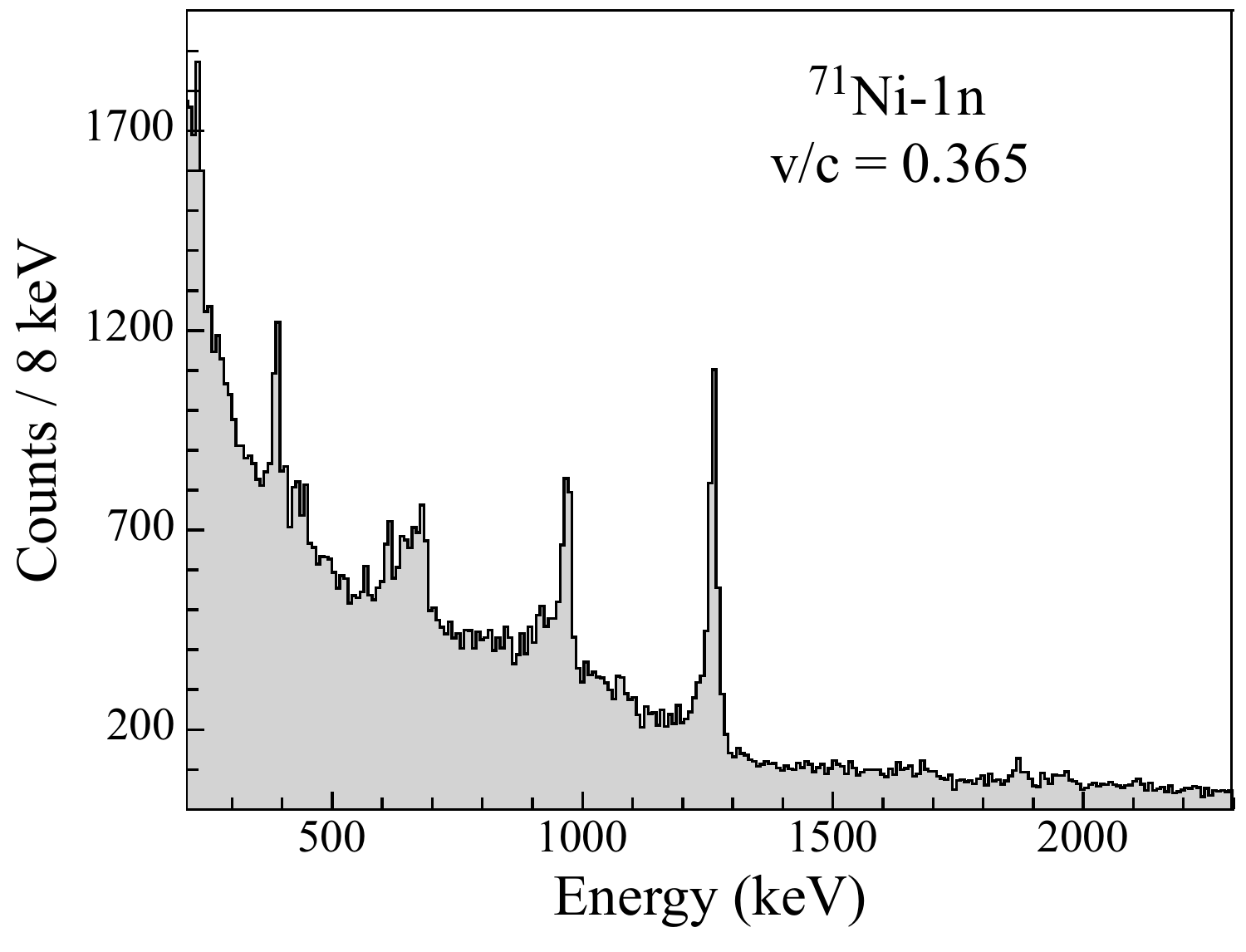}
        \caption{ 
          Doppler-reconstructed $\gamma$-ray spectrum in
          coincidence with the \nuc{70}{Ni} knockout residues for the
          one-neutron knockout from \nuc{71}{Ni}. 
        }
        \label{fig:specni71}
      \end{center}
    \end{figure}

    The Doppler-corrected $\gamma$-ray spectrum observed for the one-neutron
    knockout reaction can be found in Fig.~\ref{fig:specni71}.  This channel
    exhibits a considerably different final-state population pattern than that
    observed in both proton knockout reactions. The largest difference observed
    between the one-neutron and one-proton channels are a decrease in the
    feeding-subtracted population of states that may be attributed to the
    proposed prolate structure, and an increase in both that of the previously
    observed (\state{5}{-}{}) level and the population of the state decaying
    via the 385-keV $\gamma$ ray. The feeding-subtracted state populations are
    discussed in more detail in Section~\ref{sec:disc}. Note that, unlike the
    proton-knockout reactions, the neutron-knockout one is expected to directly
    populate states of negative parity. 
    
\begin{figure}[ht!]
      \begin{center}
        \pic{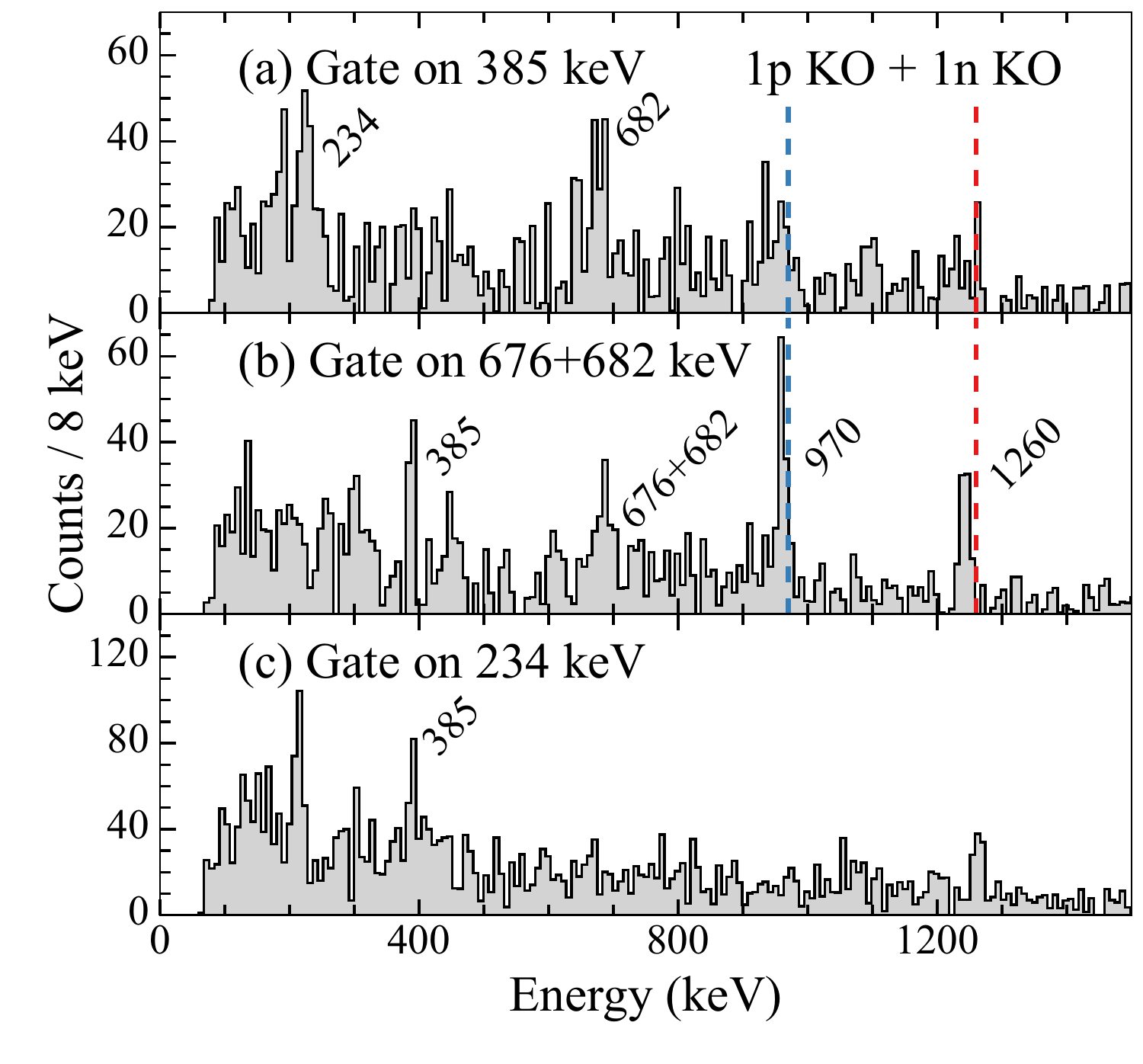}
        \caption{ 
          (Color online) Coincidence spectra for the 385-keV, the 676-682 keV doublet, and the
          234-keV transitions from the sum of the one-proton and one-neutron
          knockout reaction channels. Due to low statistics, the placement of
          the 385-keV transition remains tentative. The displacement of the
          yrast energies in panel (b), as indicated by the dashed lines at the
          expected yrast energies, is discussed in the text.
        }
        \label{fig:coinc_385}
      \end{center}
    \end{figure}

    The strongest unplaced peak in the spectrum is the previously reported
    385-keV transition~\cite{chiara2015}. It is tentatively placed as
    de-exciting a 3297-keV state, based on the coincidence relationships with
    the 682- and 234-keV transitions from the (\state{5}{-}{}) state.
    Fig.~\ref{fig:coinc_385} provides coincidence spectra resulting from the
    sum of the data for the one-proton and one-neutron reaction channels. The
    676-keV transition, observed in coincidence with the 682-keV $\gamma$ ray
    depopulating the (\state{5}{-}{}) level in the one-proton knockout
    reaction, is not observed in one-neutron knockout. As the 385-keV line
    is only weakly populated in the low-statistics two-proton knockout channel,
    data from the latter reaction were not included in the sum. Relatively weak
    coincidence relationships between the 385-, 676-682-, and the 234-keV
    transitions can be observed (see Fig.~\ref{fig:coinc_385}). 

    Aside from the limited statistics in the coincidence spectra, the position
    of the two peaks near 957 and 1249 keV in coincidence with the 676-682-keV
    doublet (Fig.~\ref{fig:coinc_385}(b)) also complicated the placement of the
    385-keV transition. As shown by the dashed lines in the figure, both yrast
    transitions appear to be offset by roughly 13 keV from their expected
    energies of 970 and 1260 keV, respectively. It is possible that this shift
    also accounts for the low-energy tails visible for these two $\gamma$ rays
    in the Doppler-reconstructed spectrum of Fig.~\ref{fig:specni71}. Energy
    shifts of this type might be due to lifetime effects and some have been
    observed in previous measurements; e.g., see Ref.~\cite{gade2019}. 
    To explore this possibility further, a simulation was carried out for a cascade
    starting from the 2912-keV, (\state{5}{-}{}) level and proceeding through
    the \state{4}{+}{1} and \state{2}{+}{1} yrast states. The (\state{5}{-}{})
    level lifetime was varied until energy shifts of the proper magnitude were
    obtained. To account for the data, a lifetime $\tau (5^-) \geq~75~\text{ps}$
    is required.

    Displacements in Doppler-corrected energy due to the lifetime of the
    decaying state can occur both because of the uncertainty in the
    determination of the particle velocity at the time of $\gamma$-ray emission
    and because the $\gamma$ decay occurs behind the target. In the former case,
    long-lived states de-excite at a lower average velocity than prompt
    transitions, as the nucleus has traversed more of the target material and
    slowed down relative to prompt mid-target emission. In the latter, the
    exact emission angle is underestimated as the decay is assumed to take
    place at mid-target, herewith resulting in a lower transition energy for
    detectors located at forward angles.
    
    Note that such displacements caused by lifetime effects could potentially
    account for the presence -- or for the partial intensity -- of some of the
    unplaced transitions listed at the bottom of Table~\ref{tab:res}. For
    example, as already pointed out above, the 957-keV $\gamma$ ray could
    originate from the improper Doppler correction of the 970-keV transition.
    Also, the 1249-keV line was placed within the proposed sequence of prolate
    levels based on observed coincidence relationships in one-proton knockout,
    but part of its intensity could conceivably be attributed to the
    contribution from a long-lived feeding state impacting the 1260-keV
    transition. Unfortunately, the level of statistics in the one-neutron
    knockout channel proved insufficient to verify the expected coincidence
    relationships within the 1249-1260 keV cascade. Furthermore, some of the
    unplaced $\gamma$ rays of Table~\ref{tab:res} could be associated with
    transitions reported in other works, for which a placement in the level
    scheme of Fig.~\ref{fig:lvlscheme} could not be confirmed because of either
    the presence of isomeric states or the lack of statistics for the
    one-neutron knockout data. For example, this may be the case for the
    915-keV line seen in the present work. It could possibly correspond to the
    transition of the same energy placed earlier~\cite{chiara2015, prokop2015}
    as feeding into the \state{6}{+}{1} level, but the anticipated coincidence
    relationships would be obscured by the $\tau =
    1.51(4)~\text{ns}$~\cite{mach2003} lifetime affecting the measured
    transition energies as well as by poor statistics.  Feeding of the
    \state{8}{+}{1} level, with its even longer lifetime of  $\tau =
    335(1)~\text{ns}$~\cite{NNDC}, could not be readily identified in the present
    measurements. Although the 183-keV $\gamma$ ray de-exciting the isomer and
    the subsequent cascade were observed in the hodoscope,
    statistics proved insufficient to connect the decay of the long-lived state
    at the focal plane with prompt transitions detected by GRETINA at the
    target.

    \begin{figure}[ht!]
      \begin{center}
        \pic[0.9]{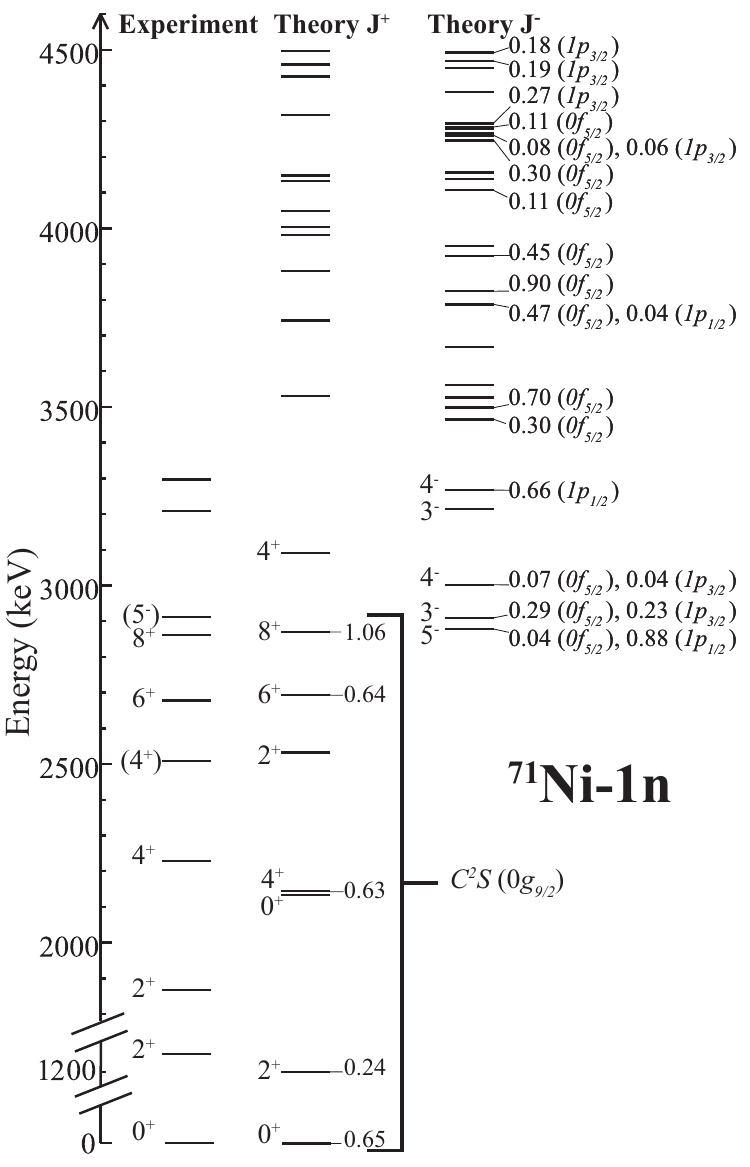}
        \caption{ 
          Comparison of the  levels populated in the one-neutron knockout
          reaction channel with results of shell-model calculations allowing
          only neutron excitations using the jj44pna effective
          interaction~\cite{lisetskiy2004}. For the positive-parity states,
          $g_{9/2}$ shell-model spectroscopic factors $C^2S$ for one-neutron
          knockout from the \nuc{71}{Ni} ground state to individual
          \nuc{70}{Ni} final states are listed when $C^2S(0g_{9/2}) > 0.1$. For
          the negative-parity levels, the $0f_{5/2},1p_{3/2}, 1p_{1/2}$
          shell-model spectroscopic factors are listed when $C^2S > 0.02$ for
          states where the sum $C^2S(0f_{5/2})+C^2S(1p_{3/2})+C^2S(1p_{1/2}) >
          0.1$. }   \label{fig:sm_comp}
      \end{center}
    \end{figure}

    Figure~\ref{fig:sm_comp} compares the \nuc{70}{Ni} levels observed solely in
    one-neutron knockout with the results of shell-model calculations carried
    out with the jj44pna effective interaction~\cite{lisetskiy2004}. The model
    space included the neutron $0f_{5/2},1p_{3/2}, 1p_{1/2},\text{ and } 0g_{9/2}$
    orbitals with the requirement that a minimum of two neutrons occupy the 
    $0g_{9/2}$ state. The calculated spectroscopic factors $C^2S$ for one-neutron
    knockout from \nuc{71}{Ni} are listed in Fig.~\ref{fig:sm_comp} for all
    levels below 4.5 MeV with $\Sigma C^2S > 0.1$. As expected, the proposed prolate
    structure is not reproduced as these calculations do not include cross-shell
    proton excitations. In addition to sizable spectroscopic strength to the
    \state{6}{+}{1} and \state{8}{+}{1} isomeric levels, it is clear from
    Fig.~\ref{fig:sm_comp} that the anticipated strength is computed to be
    fragmented among many high-energy, negative-parity \nuc{70}{Ni} states.
    Hence, it is plausible that many of the unplaced transitions in the
    one-neutron channel are associated with $\gamma$ decay from such
    negative-parity levels.

\subsection{Two-proton knockout from \nuc{72}{Zn}}
    As the \nuc{72}{Zn} beam has no isomeric component, its use in two-proton
    knockout provides a means to confirm the role of cross-shell proton
    excitations without the possibility of contamination of the measured
    populations by knockout from isomers associated with complex configurations. 
    
\begin{figure}[h]
      \begin{center}
        \pic{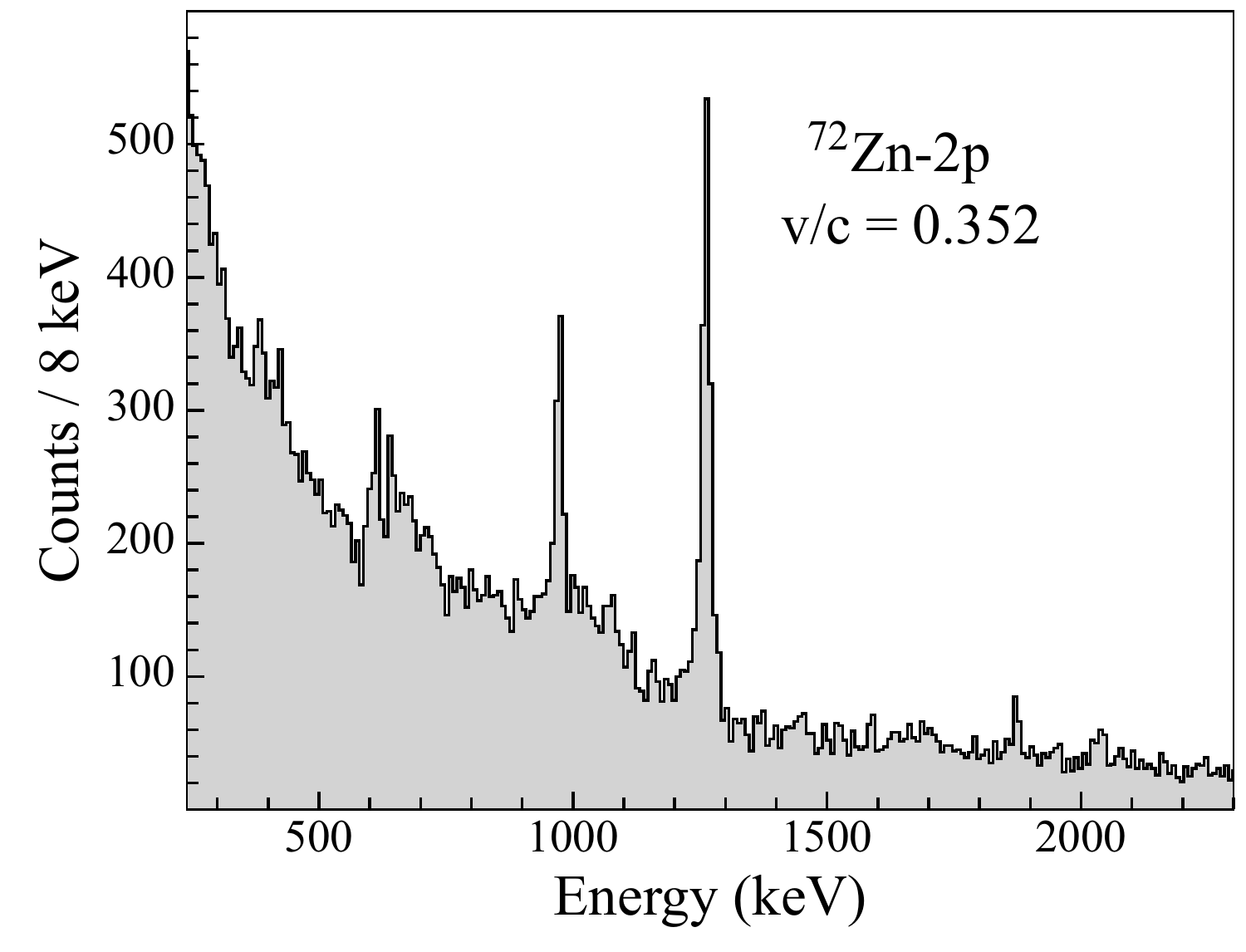}
        \caption{ 
          Doppler-reconstructed $\gamma$-ray spectrum in coincidence with the
          \nuc{70}{Ni} knockout residues for the two-proton knockout from
          \nuc{72}{Zn}.  }
        \label{fig:fitzn72}
      \end{center}
    \end{figure}

    The observed Doppler-corrected $\gamma$-ray spectrum for two-proton
    knockout populating states in \nuc{70}{Ni} is presented in
    Fig.~\ref{fig:fitzn72}. As in one-proton knockout, the two-proton knockout
    channel  predominantly populates states decaying through the yrast cascade
    to the ground state.  The strength is largely split between the yrast
    levels and the proposed prolate structure, although considerably fewer
    transitions are observed following two-proton knockout.

    As in the one-proton knockout, the 676- and 682-keV transitions 
    could not be distinguished.  Moreover, the low statistics precluded
    determining the intensity balance between the two $\gamma$ rays based on
    coincidence relationships as was done in one-proton knockout and the ratio
    determined from the latter data was relied upon. It is possible that this
    approach results in an incorrect estimate for the intensity of the 676-keV
    $\gamma$ ray as the measured intensity for the 682-keV transition combined
    with the assumption regarding the 676-682 ratio leads to an expected
    676-keV yield below the experimental sensitivity. However, assuming the
    absence of any measurable 676-keV $\gamma$ ray in the two-proton channel
    results in an increase of the 682-keV transition intensity by only $10\%$.
    In view of this small impact of this additional uncertainty on the ratio
    for the state populations, Table~\ref{tab:res} assumes that the intensity
    distribution is the same in the one- and two-proton knockout channels.  The
    adopted branching ratio from the NNDC database~\cite{NNDC} is utilized to
    determine the intensity of the 234-keV transition as the peak-to-background
    ratio in this low-energy region of the spectrum is as poor as in one-proton
    knockout.
    
    The inspection of Fig.~\ref{fig:fitzn72} also reveals the presence of 
    significant yield above 700 keV; i.e., close to the location of the 682-keV
    $\gamma$ ray. This excess of counts is seen in spectra associated with
    two-proton knockout, as a comparison of data from the different reaction
    channels in Fig.~\ref{fig:allspec_leg} indicates. The shape of this
    "structure" could possibly be viewed as corresponding to the lineshape of a
    714-keV transition emitted from a long-lived state. It could also be
    associated with a multiplet of unresolved, weak $\gamma$ rays. In any
    event, the presence of this structure affects the extraction of the
    intensity in the 682-keV peak. For example, assuming that there is a
    714-keV transition de-exciting a level with a lifetime of the order of 
    75~ps leads to a decrease in the intensity of the 682-keV $\gamma$ ray of
    $21\%$, as compared to that obtained if the structure is composed of prompt
    transitions. The errors on intensities reported in Table~\ref{tab:res} for
    the 682-keV transition and for others depending on its yield (676 and 234
    keV) reflect these difficulties.
 
\section{Discussion}
\label{sec:disc}
 
    As is discussed below, the observations reported above for the relative
    intensities of the transitions within the proposed prolate structure are
    consistent with its possible proton-hole character; i.e., with the
    association of the states involved with the predicted prolate-deformed
    structure built on the (\state{0}{+}{2}) level~\cite{tsunoda2014,prokop2015}. 

\begin{figure}[h]
      \begin{center}
        \pic{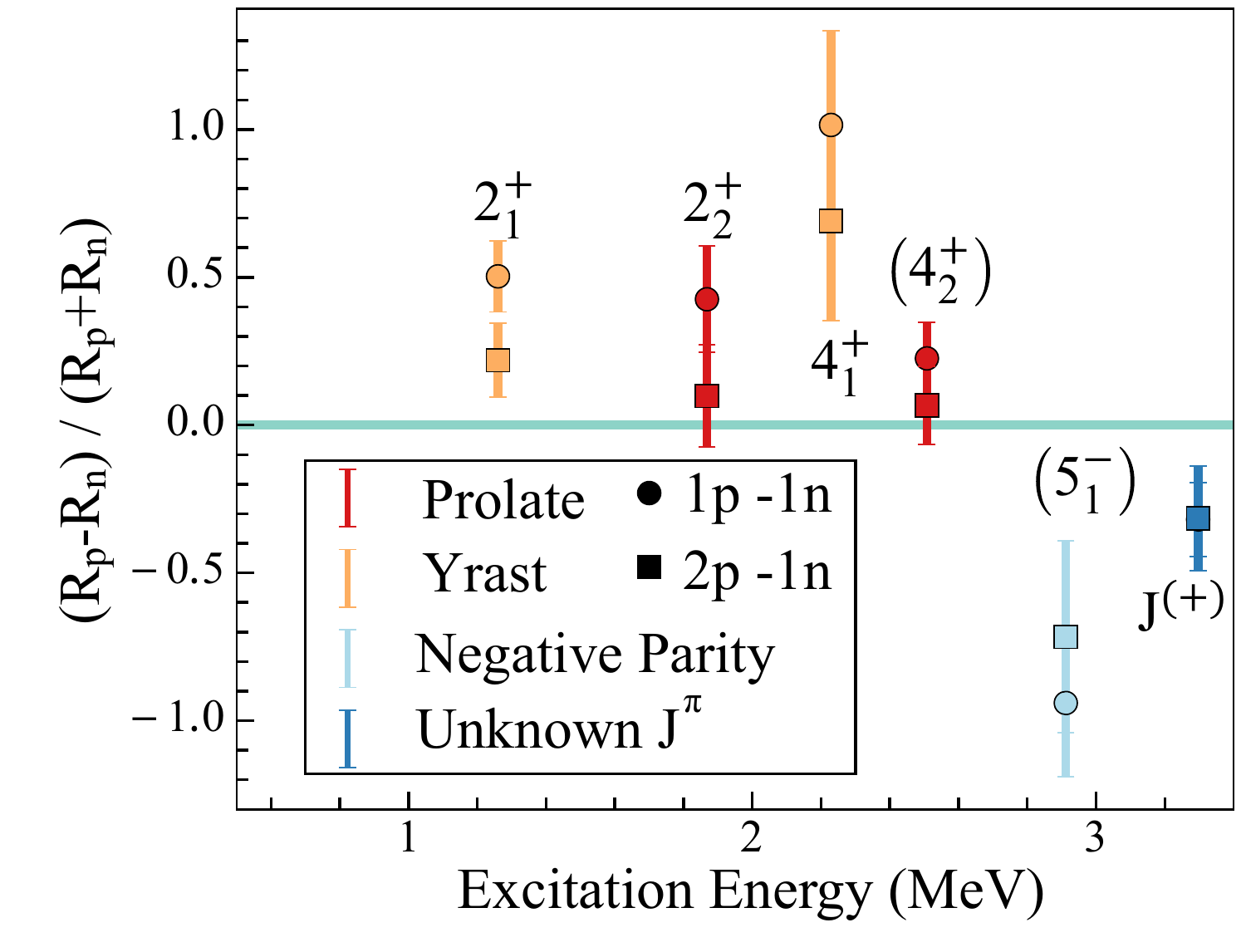}
        \caption{ 
          (Color online) Comparison of the different knockout reactions, where
          $R_p$ is the ratio of the feeding-subtracted population of a state
          from the proton-knockout reaction channels divided by the total number of
          \nuc{70}{Ni} knockout residues detected for that channel, and $R_n$
          is the same for neutrons. Plotted are differences over the sums of
          these quantities compared to the one-neutron knockout reaction for
          either the one-proton (circles) or two-proton (squares) knockout
          cases.  Points above the blue line correspond to states populated
          primarily in the proton-knockout reactions, while points below it 
          would be primarily populated in the one-neutron knockout one. The
          state of unknown spin-parity marked as $J^{(+)}$ is the 3297-keV
          level.}
      \label{fig:comp}
      \end{center}
    \end{figure}

    Differences in the properties of the states fed in the three knockout
    reactions can be inferred from Fig.~\ref{fig:comp}, where
    the intensity with which the states are produced in the various channels is
    presented as follows. First, the intensities were corrected for observed, direct
    feeding from other levels and they were then divided by the number of
    detected \nuc{70}{Ni} residues in the specific channel under consideration. 
    Figure~\ref{fig:comp} then compares the direct population of specific
    states by plotting the difference of such ratios over their sum, and
    systematically considering the proton-knockout channels versus the neutron
    one. In this approach, points with positive (negative) values are
    associated with states where larger direct population occurs via
    proton (neutron) knockout.

    Figure~\ref{fig:comp} indicates that the proposed prolate structure and the
    yrast states are populated more strongly in one- and two-proton knockout
    reactions rather than in the one-neutron knockout channel. The difference
    in population in the yrast states is likely affected to some extent by a
    combination of factors such as the incomplete subtraction of feeding of
    these states through levels of negative parity populated almost exclusively
    in neutron knockout (see discussion above) and/or the possible presence of
    long-lived states in this reaction channel, which can affect the reported
    yields. 
   
    On the other hand, the increased population of the states proposed
    to be associated with excitations built on a prolate shape is noteworthy
    and is in line with expectations based on the predicted role of proton
    cross-shell excitations in the configurations of these levels. The
    \trans{2}{+}{2}{0}{+}{2} transition was not observed in any of the present
    reaction channels. Ref.~\cite{prokop2015} places the (\state{0}{+}{2})
    state at 1567 keV, resulting in a 301-keV $\gamma$ ray linking the two
    lowest levels of the proposed prolate structure that is unlikely to compete
    with the higher-energy \trans{2}{+}{2}{0}{+}{1} (1868 keV)  and
    \trans{2}{+}{2}{2}{+}{1} (609 keV) transitions. Hence, the lack of
    observation does not appear to invalidate the proposed picture.

    A similar reasoning suggests also that the stronger population of the
    (\state{5}{-}{}), 2912-keV level in one-neutron knockout is consistent with
    the expectation that the negative-parity states are associated with neutron
    configurations at low excitation energy in \nuc{70}{Ni}. The placement of
    the 385-keV $\gamma$ ray as a transition feeding the (\state{5}{-}{})
    state contradicts the work of Ref.~\cite{chiara2015}, where this transition
    was proposed to be associated with the prolate structure.  In addition,
    Ref.~\cite{chiara2015} had also reported a 676-keV line and had speculated
    that it could be part of the proposed prolate structure based on its observation
    in the two-neutron knockout reaction channel but absence in the deep-inelastic
    scattering data. The situation is more complex here: a 676-keV line is
    present in the one-proton knockout spectrum where it is in coincidence with
    the 682-keV transition, but it is not directly observed in one-neutron
    removal.  However, as discussed above, this 676-keV energy would be
    compatible with that of a Doppler-reconstructed, 682-keV transition from a
    long-lived ($\tau \geq 75$ ps) state. These observations together with the
    noted coincidence relationships with the 682-keV,
    \transtentleft{5}{-}{}{4}{+}{1} transition makes an association with the
    proposed proton excitations unlikely.

\section{Summary}
    The structure of the \nuc{70}{Ni} nucleus has been investigated following
    one-neutron and one- and two-proton knockout reactions. A number of new
    transitions have been added to the level scheme. Furthermore, the
    population of the observed levels has been found to depend on the reaction
    channel. Specifically, preferential population in proton knockout has been
    observed for a set of states proposed in earlier work to be associated with
    a prolate-deformed structure. The finding of preferential excitation
    through one- and two-proton knockout is consistent with shell-model
    calculations that associate this structure with proton excitations across
    the $Z = 28$ shell gap.
   
\begin{acknowledgments}
This work was supported by the National Science Foundation under Grants No.
PHY-1565546 and PHY-1811855, and by the Department of Energy, Office of Nuclear
Physics, under Grant No's DE-AC02-06CH11357 (ANL), DE-FG02-94ER40834 (Maryland), DE-FG02-08ER41556
(MSU), DE-FG02-97ER41041 (UNC),  and DE-FG02-97ER41033 (TUNL).  This material is
based upon work supported by the Department of Energy National Nuclear Security
Administration through the Nuclear Science and Security Consortium under Award
Number DE-NA0003180.  GRETINA was funded by the DOE, Office of Science.
Operation of the array at NSCL was supported by DOE under Grant No.
DE-SC0014537 (NSCL) and DE-AC02-05CH11231 (LBNL).
\end{acknowledgments}

\end{document}